\documentclass[12pt]{article}
\usepackage[latin9]{inputenc}
\usepackage{float}
\usepackage{textcomp}
\usepackage{amstext}
\usepackage{graphicx}
\usepackage{esint}
\PassOptionsToPackage{normalem}{ulem}
\usepackage{ulem}

\makeatletter
\@ifundefined{date}{}{\date{}}

\usepackage{times}

\usepackage[left,modulo]{lineno}
\linenumbers

\newenvironment{sciabstract}{%
\begin{quote} \bf}{\end{quote}}

\newcounter{lastnote}

\title{
Beating Rayleigh's Curse by Imaging Using Phase Information}

\author
{Weng-Kian Tham,$^{1, 2\ast}$ Hugo Ferretti,$^{1, 2\ast}$ Aephraim M. Steinberg$^{1, 2}$\\
\\
\normalsize{$^{1}$Centre for Quantum Information and Quantum Control and Institute for Optical Sciences,}\\ 
\normalsize{Department of Physics, University of Toronto,}\\
\normalsize{60 St. George St, Toronto, Ontario, Canada, M5S 1A7}\\
\normalsize{$^{2}$Canadian Institute For Advanced Research,}\\
\normalsize{180 Dundas St. W., Toronto, Ontario, Canada, M5G 1Z8}\\
\\
\normalsize{$^\ast$ These authors contributed equally to this work.}
}

\makeatother

\begin{document}
\baselineskip24pt

\maketitle

\begin{sciabstract} Any imaging device such as a microscope or telescope
has a resolution limit, a minimum separation it can resolve between
two objects or sources; this limit is typically defined by ``Rayleigh's
criterion'' \cite{Rayleigh1879}, although in recent years there have
been a number of high-profile techniques demonstrating that Rayleigh's
limit can be surpassed under particular sets of conditions \cite{hell1994breaking,betzig2006imaging,hess2006ultra}.
Quantum information and quantum metrology have given us new ways to
approach measurement \cite{giovannetti2004quantum,giovannetti2011advances,mitchell,walther2004broglie,nagata2007beating};
a new proposal inspired by these ideas \cite{tsang2015quantum} has
now re-examined the problem of trying to estimate the separation between
two poorly resolved point sources. The ``Fisher information'' \cite{casella2002statistical}
provides the inverse of the Crámer-Rao bound, the lowest variance
achievable for an unbiased estimator. For a given imaging system and
a fixed number of collected photons, Tsang, Nair and Lu observed that
the Fisher information carried by the \emph{intensity} of the light
in the image-plane (the only information available to traditional
techniques, including previous super-resolution approaches) falls
to zero as the separation between the sources decreases; this is known
as ``Rayleigh's Curse.'' On the other hand, when they calculated the
quantum Fisher information \cite{helstrom1969quantum,petz2011introduction}
of the full electromagnetic field (including amplitude \textit{and}
phase information), they found it remains constant. In other words,
there is infinitely more information available about the separation
of the sources in the phase of the field than in the intensity alone.
Here we implement a proof-of-principle system which makes use of the
phase information, and demonstrate a greatly improved ability to estimate
the distance between a pair of closely-separated sources, and immunity
to Rayleigh's curse.

\end{sciabstract}

As an electromagnetic wave, light is characterized by both an amplitude
and a phase. Traditional imaging systems use lenses or mirrors to
refocus this wave and project an image of the source onto a screen
or camera, where the intensity (or rate of photon arrivals) is recorded
at each position. (We refer to all such techniques as ``image-plane
counting'' or IPC.) Although the phase of the wave at the position
of the optics plays a central role during the focusing, any information
about the phase \emph{in the image plane} is discarded. When light
passes through finite-sized optical elements, diffraction smears out
the spatial distribution of photons so that point-sources map (via
the point-spread-function or PSF) onto finite-sized spots at the image-plane.
Thus, our ability to resolve the point-sources is inhibited when their
separation in the image-plane, $\delta$, is comparable to or less
than the width $\sigma$ of the PSF.

The typical response to diffraction limits has been to build larger
(or higher numerical-aperture) optics, thereby making the PSF sharper/narrower.
In recent years, techniques have been developed in specific cases
that address these limits in more novel ways \cite{hell1994breaking,betzig2006imaging,hess2006ultra,hemmer2012universal,giovannetti2009sub,tsang2009quantum,PhysRevLett.107.083603,PhysRevLett.112.223602,PhysRevLett.97.163903,schwartz2013superresolution}.
Despite their success, these techniques require careful control of
the source of illumination, which is not always possible in every
imaging application (e.g. astronomy). In order to beat the diffraction
limit for \emph{fixed, mutually incoherent} sources, a paradigm shift
arising from the realisation that there is a huge amount of information
available in the phase discarded by IPC may prove revolutionary.

It was shown in \cite{tsang2015quantum} that whereas in IPC the Fisher
Information, $I_{f}$, vanishes quadratically with the separation
$\delta$ between two equal-intensity incoherent point sources of
light, it remains undiminished when the full electromagnetic field
is considered. Now $I_{f}$ is related to the performance of a statistical
estimator by : 
\begin{equation}
\mbox{Var}\left(\delta_{est}\right)\geq\frac{1}{I_{f}}\left(1+\frac{\partial\left(\mbox{bias}\right)}{\partial\delta_{actual}}\right)\label{eq:CRLB}
\end{equation}
where $\delta_{est}$ is some estimator of $\delta_{actual}$ and
$\mbox{bias}\equiv\left\langle \delta_{est}\right\rangle -\delta_{actual}$.
The vanishing of $I_{f}$ as $\delta\rightarrow0$ suggests that for
closely separated sources, the variance in an IPC-based estimate of
$\delta$ is cursed to blow up. Its independence of $\delta$ for
the full field, on the other hand, immediately leads to the idea that
this divergence can be averted by using phase as well as intensity
information.

One natural way to do this would be to use SPAtial mode DEmultiplexing
(SPADE)\cite{tsang2015quantum,richardson2013space}, in which incoming
light is decomposed into its Hermite-Gauss (HG) components and the
amplitude of each is measured. It can be shown that the full set of
HG amplitudes contains the same $I_{f}$ as the full EM field. A reduced
version called binary SPADE prescribes discriminating only between
the $TEM_{00}$ mode and the sum of all other modes. For small $\delta$,
only one other mode acquires significant amplitude in any case, so
the $I_{f}$ available to binary SPADE becomes essentially equal to
the full Fisher information. The method can be understood as follows:
the projection always succeeds ($P_{00}=\left|\left\langle \psi|TEM_{00}\right\rangle \right|^{2}=1$)
when the two point-sources are overlapped ($\delta=0$), but has a
failure probability $1-P_{00}$ which grows quadratically with $\delta$
\cite{NoteTEM00vsAiry}. Knowing the $TEM_{00}$ component as a proportion
of all HG amplitudes (i.e. $P_{00}$ \emph{and} $1-P_{00}$) allows
one to deduce $\delta$.

Experimentally however, merely capturing the $TEM_{00}$ component
(say, by coupling into a single mode fiber) without a normalization
factor (which allows us to deduce $1-P_{00}$) provides no advantage
over IPC. Practically speaking, the crucial information comes from
a projection onto some mode {\em orthogonal} to $TEM_{00}$ in
order to estimate $1-P_{00}$. While a mode such as $TEM_{10}$ would
contain all the information (for a separationin the x-direction in
that example), the same scaling can be obtained by projecting onto
any spatially antisymmetric field mode. As a proof-of-principle, we
have designed and implemented an experimentally convenient method,
SPLICE (Super-resolved Position Localisation by Inversion of Coherence
along an Edge), which instead carries out one technically straightforward
projection. Consider the mode function $\psi_{\perp}\left(x,y\right)=\exp\left(-\frac{x^{2}+y^{2}}{4\sigma^{2}}\right)\mbox{sign}\left(x\right)$,
constructed such that its inner product with $TEM_{00}$ vanishes.
The probability that such a projection succeeds is: 
\begin{equation}
P_{\perp}=\frac{1}{2}\left(\left|\left\langle \psi_{1}|\psi_{\perp}\right\rangle \right|^{2}+\left|\left\langle \psi_{2}|\psi_{\perp}\right\rangle \right|^{2}\right)=e^{-2\Delta}\text{erf}^{2}\sqrt{\Delta}\label{eq:-1}
\end{equation}
where $\Delta=\delta^{2}/32\sigma^{2}$, and $\delta$ is the separation
between point sources on the image plane, and $\psi_{1/2}$ is the
field from each source.

The per photon Fisher information can be written as 
\begin{eqnarray}
I_{f} & = & \frac{\left(e^{-\Delta}\sqrt{\pi\Delta}\text{erf}\sqrt{\Delta}-e^{-2\Delta}\right)^{2}}{2\pi\sigma^{2}}+\frac{\left(e^{-\Delta}\sqrt{\pi\Delta}\text{erf}^{2}\left(\sqrt{\Delta}\right)-e^{-2\Delta}\text{erf}\sqrt{\Delta}\right)^{2}}{2\pi\sigma^{2}\left(e^{2\Delta}-\text{erf}^{2}\sqrt{\Delta}\right)}\label{eq:-2}
\end{eqnarray}
where the first term comes from $P_{\perp}$ and the second from $1-P_{\perp}$.
Crucially, as $\Delta\to0$, $1-P_{\perp}$ vanishes, meaning that
an experimentally simple scheme for projecting only onto $\psi_{\perp}$
does as well as a more complicated scheme which could measure multiple
projections simultaneously. In (\textit{Fig. 1.}), we plotted the
Fisher information for SPLICE in comparison with other methods. It
is easy to see that it remains constant as $\delta\rightarrow0$,
evading Rayleigh's curse, and extracting nearly $2/3$ of the total
information available to full SPADE using an experimentally simple
technique. More sophisticated methods relying on waveguides or cavities
could be designed to approach 100\% of the total $I_{f}$.
\begin{figure}[H]
\begin{centering}
\includegraphics[width=13cm]{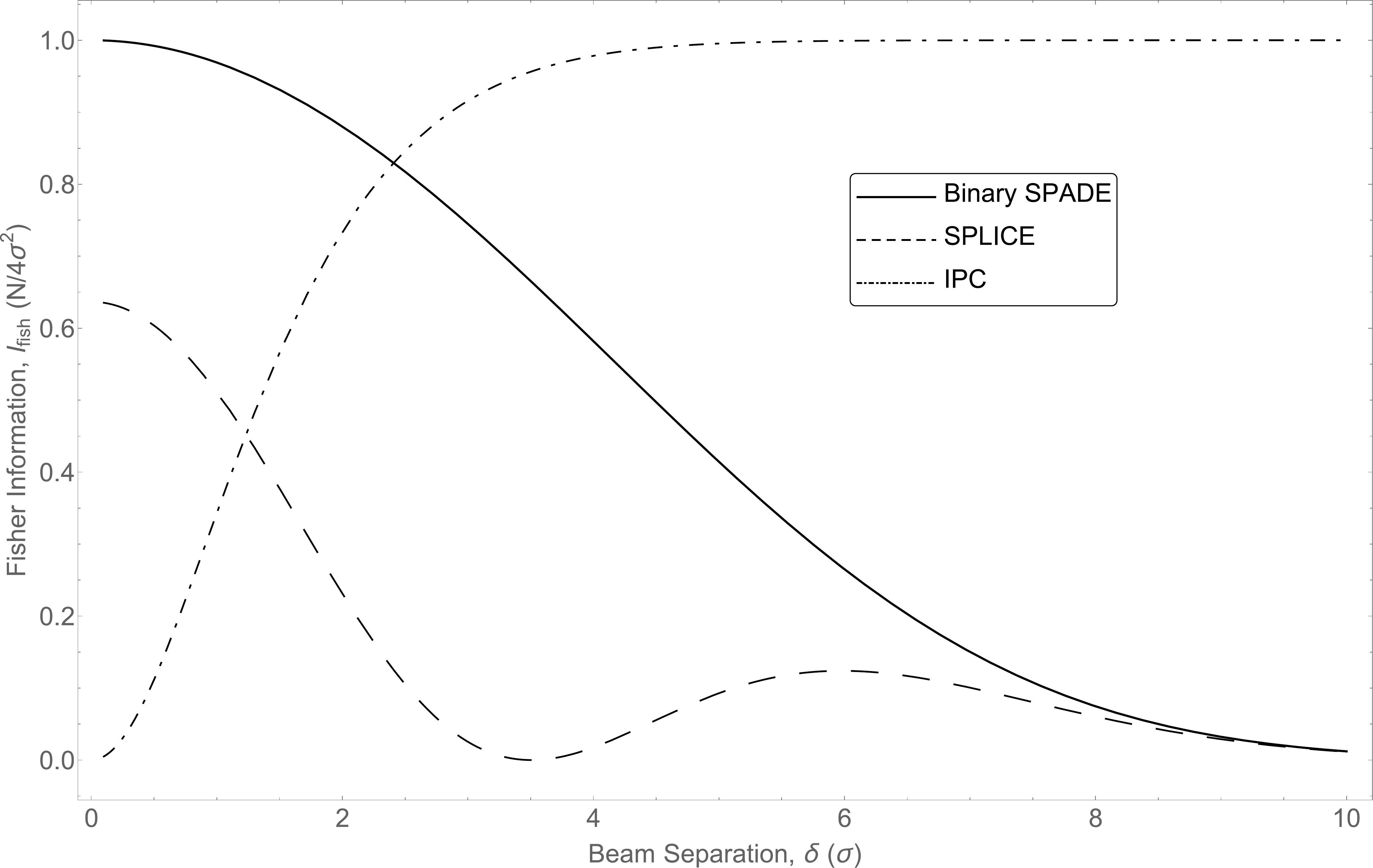}
\par\end{centering}

\caption{\textbf{Fisher information for SPLICE, binary SPADE, and IPC} Theory
plot of Fisher information for various methods vs beam separation
$\delta$, normalized to units of $N/4\sigma^{2}$ and $\sigma$ respectively.}
\end{figure}

In order to experimentally demonstrate improved performance over IPC,
we used two mutually incoherent collimated $TEM_{00}$ Gaussian beams
in place of distant point sources and an imaging optical setup. The
beams were directed through a Sagnac-like beam displacer shown in
(\textit{Fig. 2.}). By moving a mirror on a motorized translation
stage as shown, we precisely control the separation $\delta$ between
the otherwise parallel beams. The separation is induced symmetrically,
such that the geometrical centroid $\left(x_{0},y_{0}\right)$ remains
static.

Our light source is an $805$-nm heralded single-photon source which
relies on type-I spontaneous parametric down-conversion (SPDC) in
a 2mm-thick BBO crystal. The crystal is pumped by $402.5$ nm light
obtained from a frequency-doubled 100-fs Ti:Sapph laser. One photon
from the SPDC pair is used to herald the presence of a signal photon
and as a means of rejecting spurious background light (our accidental
coincidences average $2\pm1$counts/sec). In order to emulate two
point sources, the other photon is split at a 50/50 fiber-splitter
and out-coupled to free-space. The two resulting beams are incoherent;
they have splitter-to-coupler distances that differ by 5cm whereas
the SPDC photons are filtered to $\Delta\lambda=3nm$ (i.e. coherence
length $\approx10\mu m$). ND filters were used to reduce the intensity
imbalance between the beams to $\approx\left(3\pm3\right)\%$.
\begin{figure}[H]
\begin{centering}
\includegraphics[width=13cm]{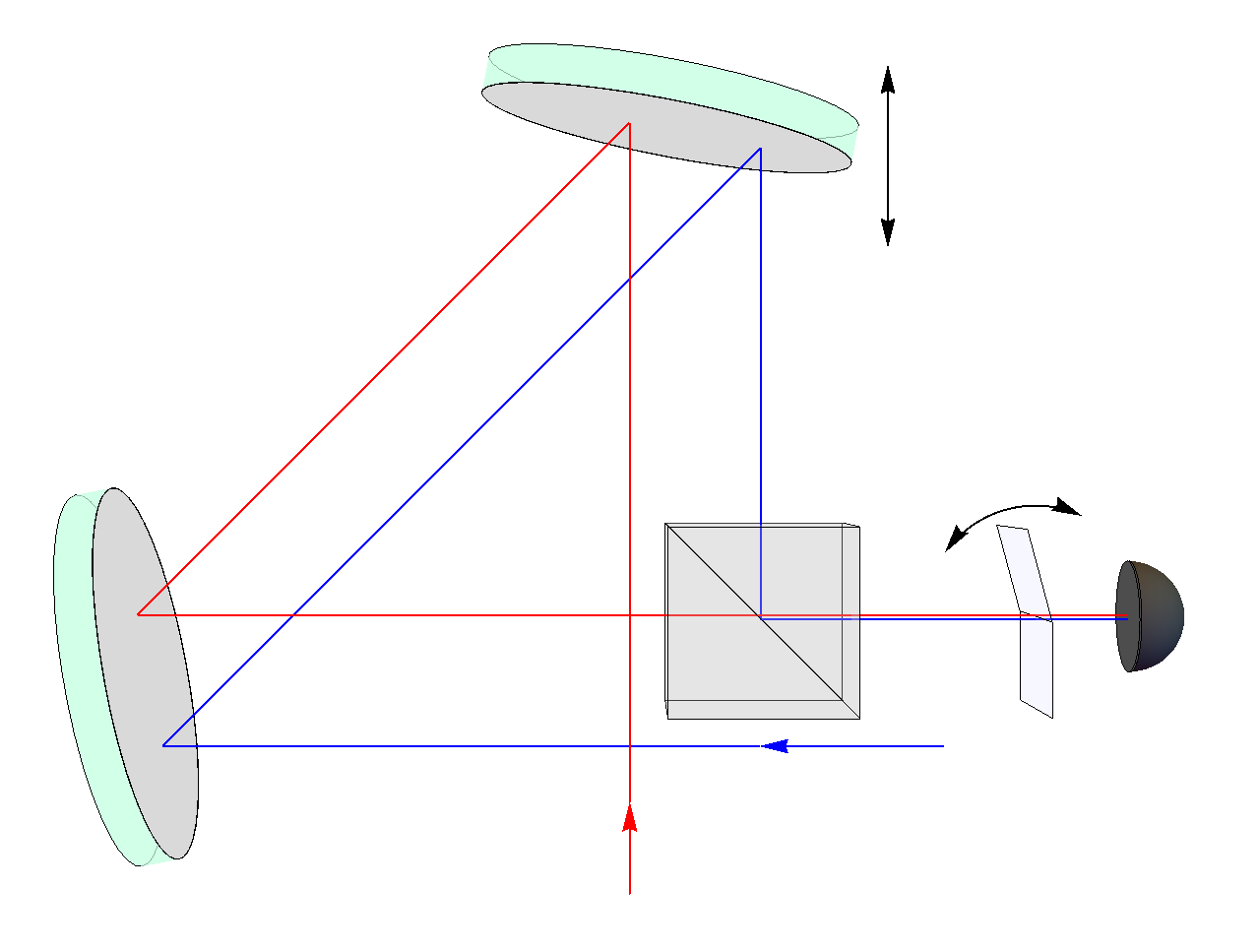}
\par\end{centering}

\caption{\textbf{Cartoon of experiment}. Shown is the apparatus for generating
the symmetrically displaced pair of sources, and the phase-shifter
used for implementing SPLICE.}
\end{figure}

At zero separation, the beams are overlapped and are both coupled
into single-mode $TEM_{00}$ fiber (coupling efficiencies are $90\%$
and $85\%$ respectively). Collimation of the beams is such that their
waists are closely matched immediately before the fiber coupler ($\sigma=614\pm4\mu m$
and $594\pm10\mu m$) in order to emulate Gaussian point spread functions
of distant sources. The projection onto $\psi_{\perp}$ is achieved
by inserting a phase plate immediately in front of the coupler such
that when $\delta=0$, a semi-circular cross-section of the beams
undergoes a $\pi$-phase shift whereas the other half experiences
none. The phase plate consists of a transparent glass microscope slide
with a sharp rectangular edge mounted on a translation and rotation
stage and positioned/rotated to minimize coupling into an otherwise
well-aligned coupler. Typical visibility is $\geq99\%$.

To compare the performance of our method with a more traditional imaging
setup, we replaced the phase plate with a 200\textmu m slit that served
as the image plane, coupling all the light transmitted through the
slit into a multimode fibre. Scanning the slit, we were able to perform
one-dimensional IPC.

With SPLICE,\emph{ }the separation of the incoherent beams was scanned,
with the detectors counting for 1 second at each step. Two sets of
these ``phase-plate'' scans were performed, one at coarse intervals
of $\delta$ (spanning $-1.96mm$ to $+1.94mm$, in steps of $0.1mm$)
and another at finer intervals ($-0.56mm\leq\delta\leq+0.44mm$ in
steps of $0.04mm$). Data from nine repetitions of the coarse scan
and fifteen of the fine scan were\emph{ }recorded.
\begin{figure}[H]
\begin{centering}
\includegraphics[width=18cm]{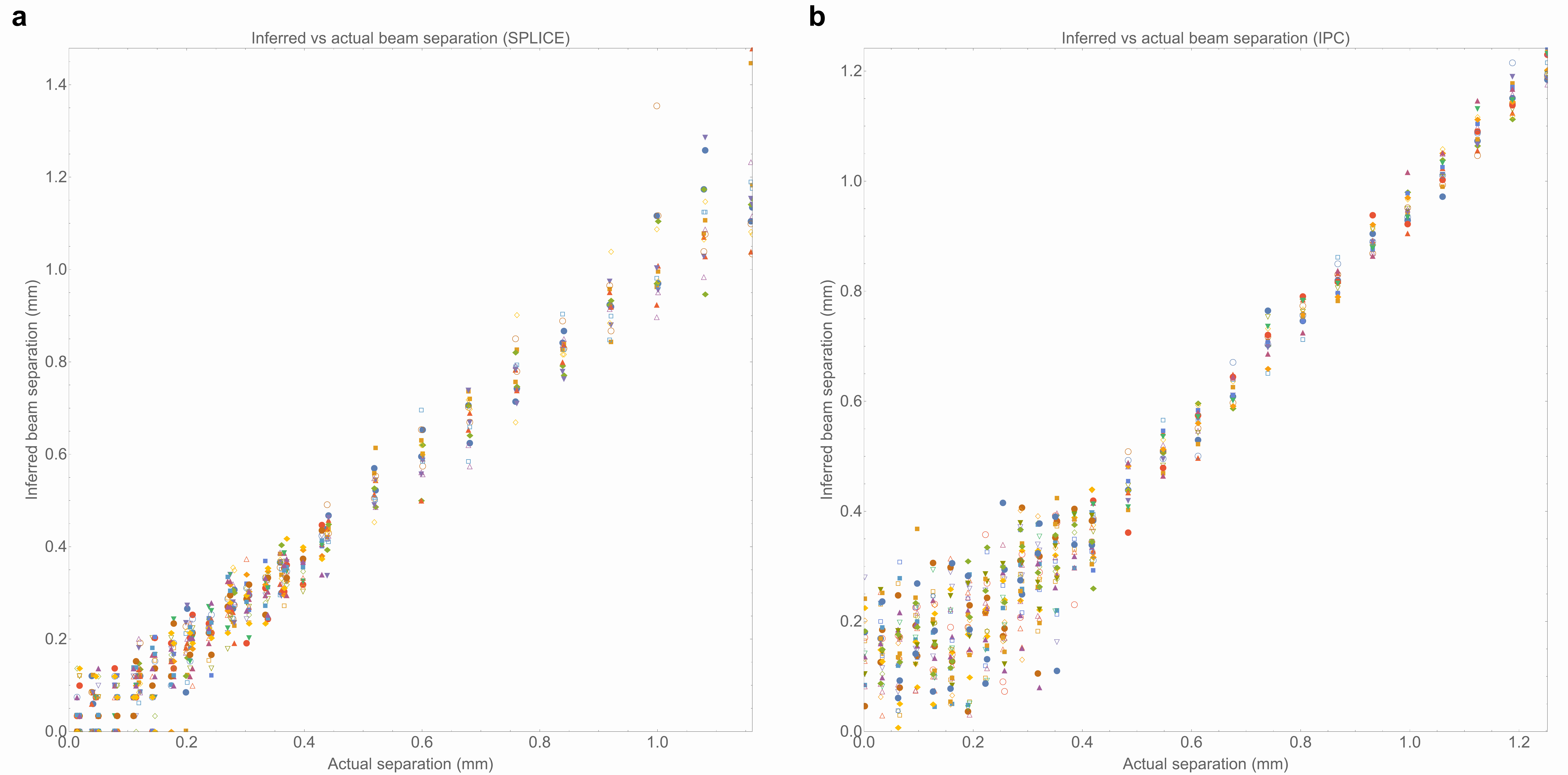}
\par\end{centering}

\caption{\textbf{Inferred separation vs known actual separation} for a) SPLICE
(from ``lookup'' on calibration curve) and b) IPC.}
\end{figure}

Whereas the ideal functional form for the resulting counts vs separation
$\delta$ is proportional to equation \ref{eq:-1}, we add a constant
$\gamma$ to account for residual background counts: 
\begin{equation}
\mbox{SPLICE counts}=\alpha\,\exp\left(-\frac{\delta^{2}}{16\sigma^{2}}\right)\mbox{erf}^{2}\left(\frac{\delta}{4\sqrt{2}\sigma}\right)+\gamma\label{eq:-3}
\end{equation}
A calibration curve was obtained from a least squares fit of this
function to data from a longer run (2 sec count time instead of 1
sec for each $\delta$), from which the beam waist $\sigma$, position
offset for the motorized translation stage, and count offset $\gamma$
due to accidentals/background were deduced. This step might be thought
of as being analogous to characterizing one's imaging optics before
use. One might then use the curve as a ``lookup table'' from which
$\delta$ is estimated from count rates. We performed such a lookup
with the remainder of our data. The \emph{estimated} $\delta$ is
plotted versus their \emph{actual }values (as reported by motorized
translation stage controllers) in (\textit{Fig. 3a.}).

The traditional image plane counting data were acquired using the
configuration described above, scanning the 200\textmu m slit between
$-1mm$ and $+1mm$ of the centroid of the two beams, counting for
4 seconds at each step. Again, we repeated this for various separations
$\delta$ and in turn repeated the whole scan several times. As before,
a set of coarse scans ($-0.04mm\leq\delta\leq1.56mm$ in steps of
$0.08mm$, $16$ repetitions) and a set of fine scans ($-0.52mm\leq\delta\leq0.44mm$
in steps of $0.04mm$, $17$ repetitions) were performed Estimation
of $\delta$ in this IPC comparison was done by least-squares fitting
the resulting image plane intensity profile to 
\begin{equation}
\mbox{IPC counts}=\alpha\left[\exp_{-}+\exp_{+}+\gamma\right]\label{eq:-4}
\end{equation}
where $\exp_{\pm}=\exp\left[-\left(x\pm\delta/2\right)^{2}/2\sigma^{2}\right]$.
Again, a calibration waist $\sigma$ and background $\gamma$ were
obtained beforehand, leaving the scale $\alpha$ and separation $\delta$
as the only fitting parameters. In practice, the fitting procedure
used to fit IPC data for small $\delta$ was different from the one
used to treat data for large $\delta$. For the latter, we simply
used built-in numerical algorithms in Mathematica and NumPy. For small
$\delta$'s however, the routines exhibited convergence and stability
issues, forcing us to Taylor expand equation \ref{eq:-4} to 2nd order
in $\delta$ and manually invert the resulting $2\times2$ design
matrix. The resulting estimated separations are plotted against actual
separations in (\textit{Fig. 3b.}). As is immediately apparent, for
separations below about $0.25$mm (approximately $0.4\sigma$), the
spread of the IPC data begins to grow, while that of the SPLICE data
remains essentially constant.

Two key metrics for the performance of either method are the standard
deviation or SD (i.e. ``spread'') and root-mean-square error (RMSE)
of the estimated beam separation. The SD measures the \textit{precision}
of a dataset but not necessarily its accuracy, while the RMSE is sensitive
to the accuracy since it quantifies the error relative to a known
actual value and not simply the reported result. In (\textit{Fig.
4.} and \textit{Fig. 5.}), SD and RMSE are plotted versus known actual
separations.
\begin{figure}[H]
\begin{centering}
\includegraphics[width=13cm]{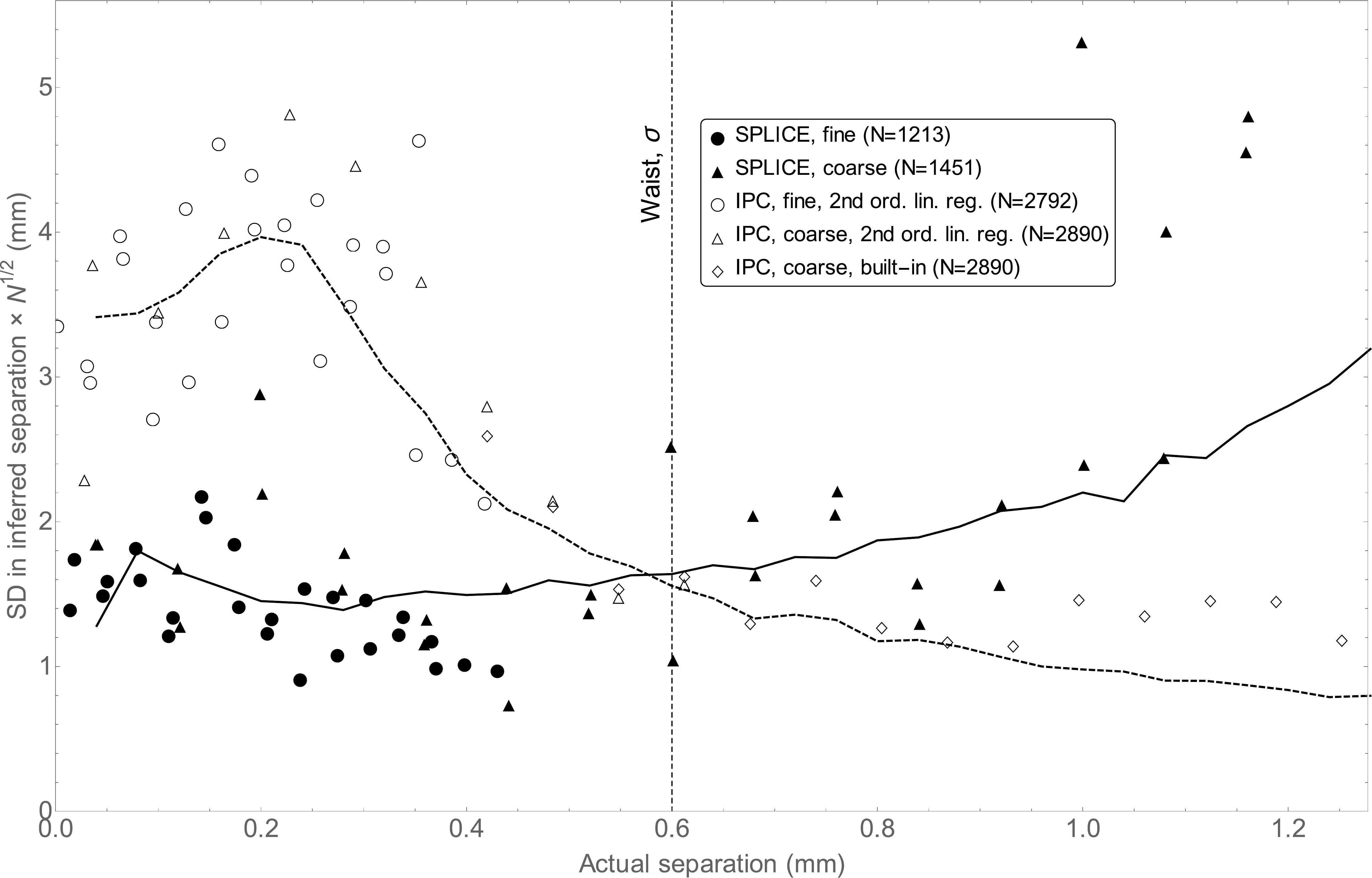}
\par\end{centering}

\caption{\textbf{Uncertainty comparison between SPLICE and IPC.} Standard Deviation
(SD) in the estimated separation plotted as a function of actual separation
for both IPC and SPLICE. The computed SD for each dataset is \emph{multiplied}
by photon number ($\sqrt{N}$). Overlayed lines are the corresponding
Monte Carlo simulations. Note that two methods were used in the fitting
of IPC data to equation \ref{eq:-4}; for small $\delta$ $\left(<0.65mm\right)$,
equation \ref{eq:-4} was expanded to 2nd order and linear regression
was performed whereas for large $\delta$ $\left(>0.4mm\right)$,
a nonlinear fitting routine built into Mathematica was used.}
\end{figure}
\begin{figure}[H]
\begin{centering}
\includegraphics[width=13cm]{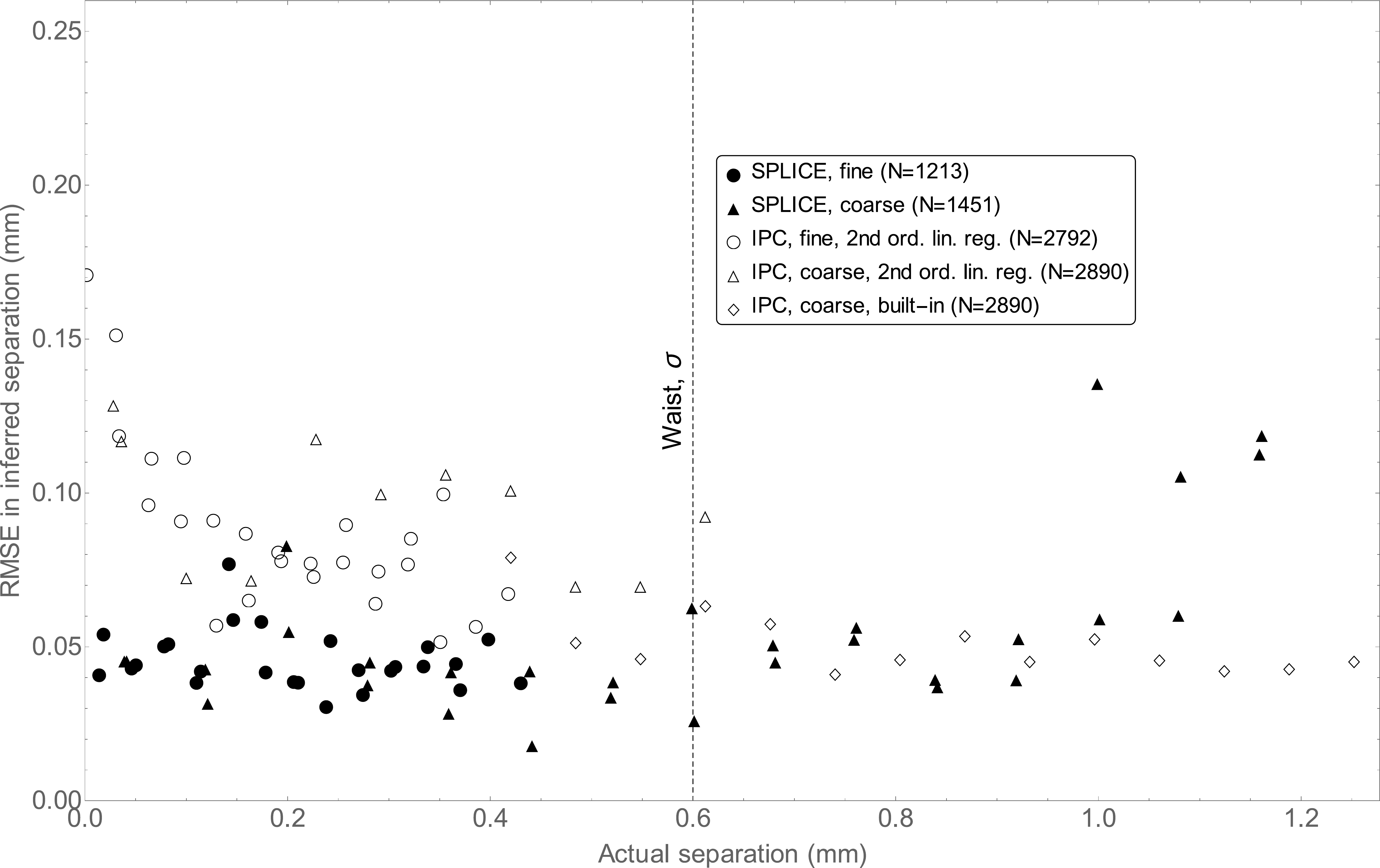}
\par\end{centering}

\caption{\textbf{Un-normalized Root Mean-Square Error (RMSE)} in the estimated
separation vs actual separation for both IPC and SPLICE. Unlike SD,
the RMSE allows us to gauge absolute error relative to the known actual
value of the parameter being estimated so that biases are accounted
for. Note that two methods were used in the fitting of IPC data to
equation \ref{eq:-4}; for small $\delta$ $\left(<0.65mm\right)$,
equation \ref{eq:-4} was expanded to 2nd order and linear regression
was performed whereas for large $\delta$ $\left(>0.4mm\right)$,
a nonlinear fitting routine built into Mathematica was used.}
\end{figure}

In order to ensure a reasonably even-footed comparison between IPC
and SPLICE, the spreads in inferred separation plotted in (\textit{Fig.
4.}) are scaled by $\sqrt{N}$ (where $N$ is the photon number that
comprises a measurement) to reflect the fact that noise in photon
counts at our detector is Poissonian. For IPC, $N$ is simply the
total photons that comprise an ``image'' on the image plane, which
in our case is actually a set of photon counts, one at each position
of the 200\textmu m slit. For SPLICE, during a calibration run, we
estimate $N$ by counting at our detector over a 1 second window while
both beams are centered (i.e. $\delta=0$) on the coupler into $TEM_{00}$
fiber with the \emph{phase-plate removed}. Since our source intensity
is stable, this gives us an \emph{estimate} of the number of incident
photons for subsequent measurements when $\delta\neq0$.

The RMSE plotted in (\textit{Fig. 5.}) is \emph{not} similarly normalized
because in addition to possible systematics, the inferred separation
is biased relative to the actual separation when $\delta$ is small.
A priori, there is no reason to suspect either bias or systematics
to scale as $\sqrt{N}$. Despite \emph{not} normalizing and despite
using approximately twice as many photons, the IPC method performs
noticeably \emph{worse} than SPLICE when $\delta<0.6mm$.

The attentive reader will note that while the spread is greater for
IPC, it does not diverge as $\delta\rightarrow0$. In fact, it would
be implausible for the uncertainty on $\delta$ to ever exceed $\sigma$.
The apparent discrepancy with the vanishing of the Fisher information
can be understood by recognizing that (as is clear from inspection
of (\textit{Fig. 1.}) at small $\delta$) the practically implemented
IPC estimator is not unbiased. To better understand the bounds on
the advantage that one can expect of SPLICE over IPC, we return to
equation \ref{eq:CRLB}. Clearly, one needs to know the bias to evaluate
the RHS. For SPLICE, the only potential source of bias is the lookup
procedure. If, for example, a less-than-perfect visibility results
in a calibration curve that does not vanish at $\delta=0$, then one
might obtain ``unphysical'' datapoints that fall \emph{under} the
minima of the calibration curve, thereby resulting in a bias when
a lookup is attempted. In our case, this is negligible since our visibility
exceeds $99\%$. The CRLB is therefore just the reciprocal of $I_{f}$,
implying a $1/\sqrt{N}$ scaling in the spread of $\delta_{est}$.

With IPC, the least-squares estimate of $\delta$ is heavily biased
at small $\delta$. An intuitive way to understand this is to note
that since the problem being addressed is the resolving of two \emph{equal
intensity }sources, the $+\delta$ and $-\delta$ cases are physically
indistinguishable. This is equivalent to saying that the power series
expansion of equation \ref{eq:-4} consists only of \emph{even }powers
of $\delta$. Given no additional information about the sign of $\delta$,
we restrict $\delta$ to be positive without loss of generality. But
in doing so, as long as spread in the estimated $\delta$ is non-zero,
the \emph{mean }estimated $\delta$ is never zero. (\textit{Fig. 6.})
shows a plot of mean inferred $\delta$ (averaged across all our datasets)
vs actual $\delta$. Overlayed is a theory curve for IPC, which takes
into account an expected bias at small $\delta$.

In \cite{suppMat}, we present theory showing that the bias term for
IPC falls to $-1$ sufficiently quickly that the RHS of inequality
\ref{eq:CRLB} tends to a finite value as $\delta\to0$. That finite
value is shown to scale as $N^{-1/4}$, which is in stark contrast
to the behaviour of the spread at large $\delta$ (for IPC) as well
as for SPLICE (at all $\delta$), where a Poissonian scaling of $N^{-1/2}$
is obeyed. This scaling is further substantiated with Monte-Carlo
simulations shown in a figure in \cite{suppMat}. Thus while SPLICE
does not offer an \emph{infinite} advantage over IPC as a naive analysis
might have us believe, it \emph{does }nevertheless offer a substantial
improvement in the absolute error and the scaling with photon number,
while simultaneously eliminating the problem of bias.
\begin{figure}[H]
\centering{}\includegraphics[width=13cm]{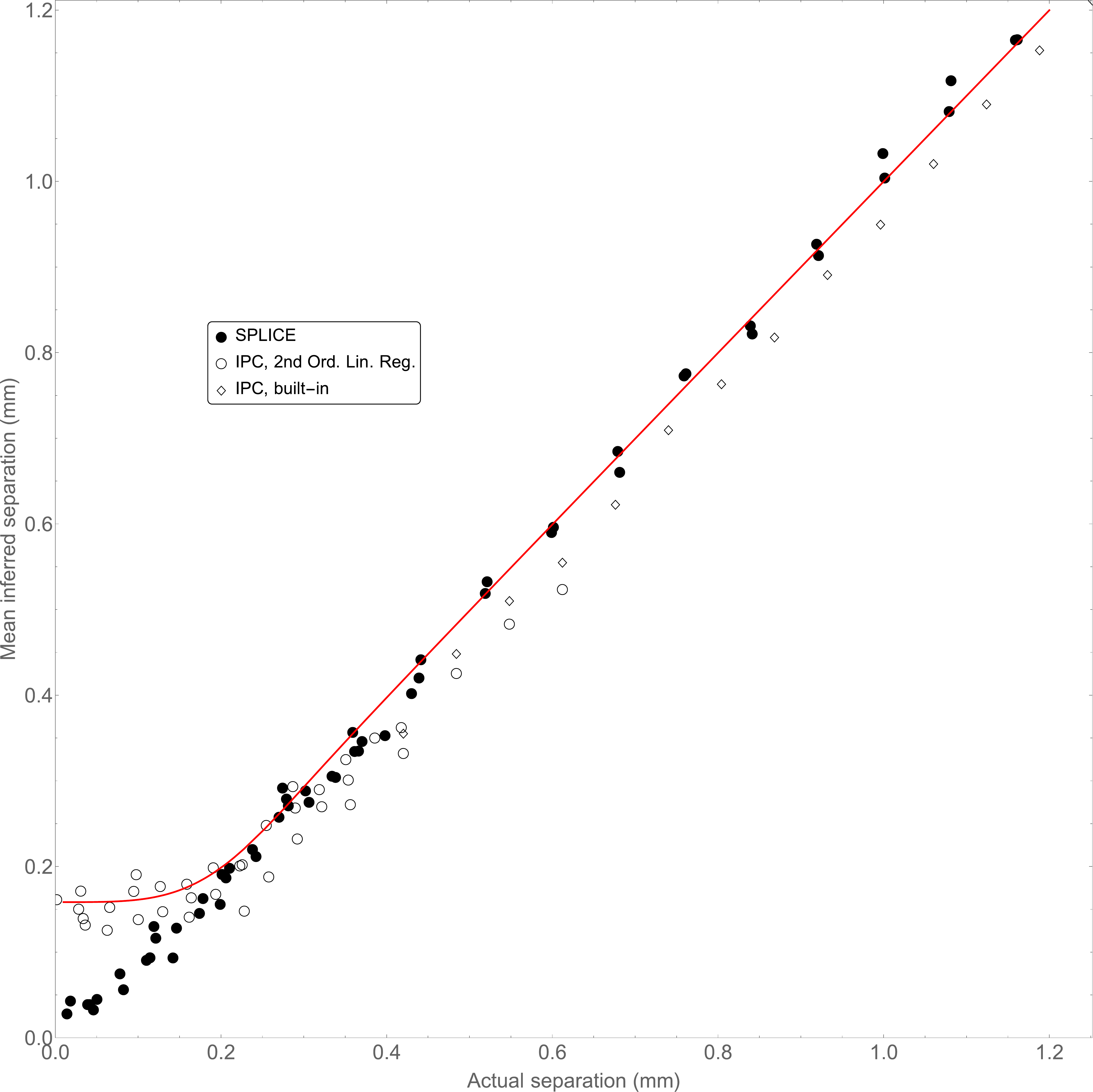}\caption{\textbf{Averaged measured separations.} Mean estimated $\delta$ for
IPC and SPLICE plotted against known actual $\delta$. Two methods
were used in the fitting of IPC data to equation \ref{eq:-4}; for
small $\delta$ $\left(<0.65mm\right)$, equation \ref{eq:-4} was
expanded to 2nd order and linear regression was performed whereas
for large $\delta$ $\left(>0.4mm\right)$, a nonlinear fitting routine
built into Mathematica was used.}
\end{figure}

In summary, we have developed and demonstrated a simple technique
that surpasses traditional imaging in its ability to resolve two closely
spaced point-sources. Furthermore, unlike existing superresolution
methods, ours requires no exotic illumination with particular coherence/quantum
properties and is applicable to classical incoherent sources. Crucially,
as a proof-of-principle, this technique highlights the importance
of realising that diffraction-imposed resolution limits are \emph{not}
a fundamental constraint but, instead, the consequence of traditional
imaging techniques discarding the phase information present in the
light. We expect that this and other related techniques that do not
discard the phase information will be developed in the future for
a broad range of imaging applications.

\emph{Note: }While preparing this manuscript, we came to realise that
similar work was being pursued by Yang et al \cite{2016arXiv160602662Y}
and Sheng et al \cite{sheng2016fault}.

\bibliographystyle{naturemag}
\bibliography{Branch-ScienceReports1}

\begin{thebibliography}{10}
\bibitem{NoteTEM00vsAiry}
 \bibinfo{note}{For circular apertures, the point spread functions are Airy
  rings. In practice, these are often well-approximated by Gaussians. Within
  this approximation, two overlapped point-sources map simply to a single
  Gaussian.}
\end{thebibliography}

\section*{Acknowledgment}

This work was funded by NSERC, CIFAR, and Northrop-Grumman Aerospace
Systems. We would like to thank Mankei Tsang for useful discussions
and the Facebook post which led to this collaboration.

\section*{Author Contribution}

W.K.T. and H.F. jointly conducted theoretical analysis, designed and
performed the experiment, and analyzed data. A.M.S. conceived of the
SPLICE scheme, guided its experimental implementation, and oversaw
the research project. All authors contributed to the writing of this
article.

\section*{Competing Interest}

We declare no competing financial interests.

\section*{Correspondence}

Correspondence and material requests should be adressed to H. F. (hferrett@physics.utoronto.ca).

\pagebreak{}

\section*{Extended Data}

\textbf{}
\begin{figure}[H]
\begin{centering}
\includegraphics[width=13cm]{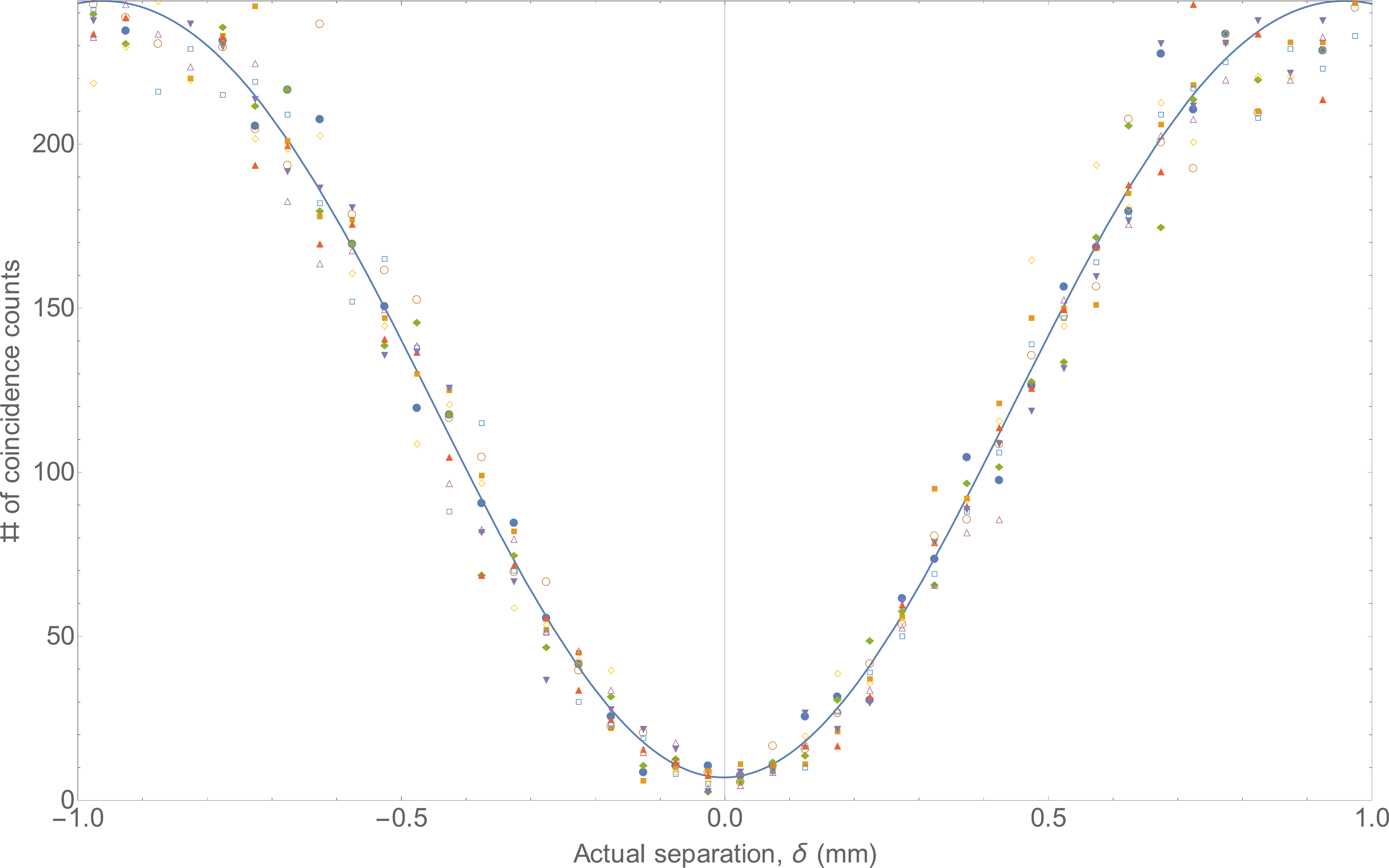}
\par\end{centering}

\textbf{\caption{\textbf{Raw data plot for SPLICE coarse scans. }Dots are experimental
photon coincidence counts plotted versus actual beam seperation $\delta$.
Solid overlay is a fit to equation \ref{eq:-1}.}
}
\end{figure}

\begin{figure}[H]
\begin{centering}
\includegraphics[width=13cm]{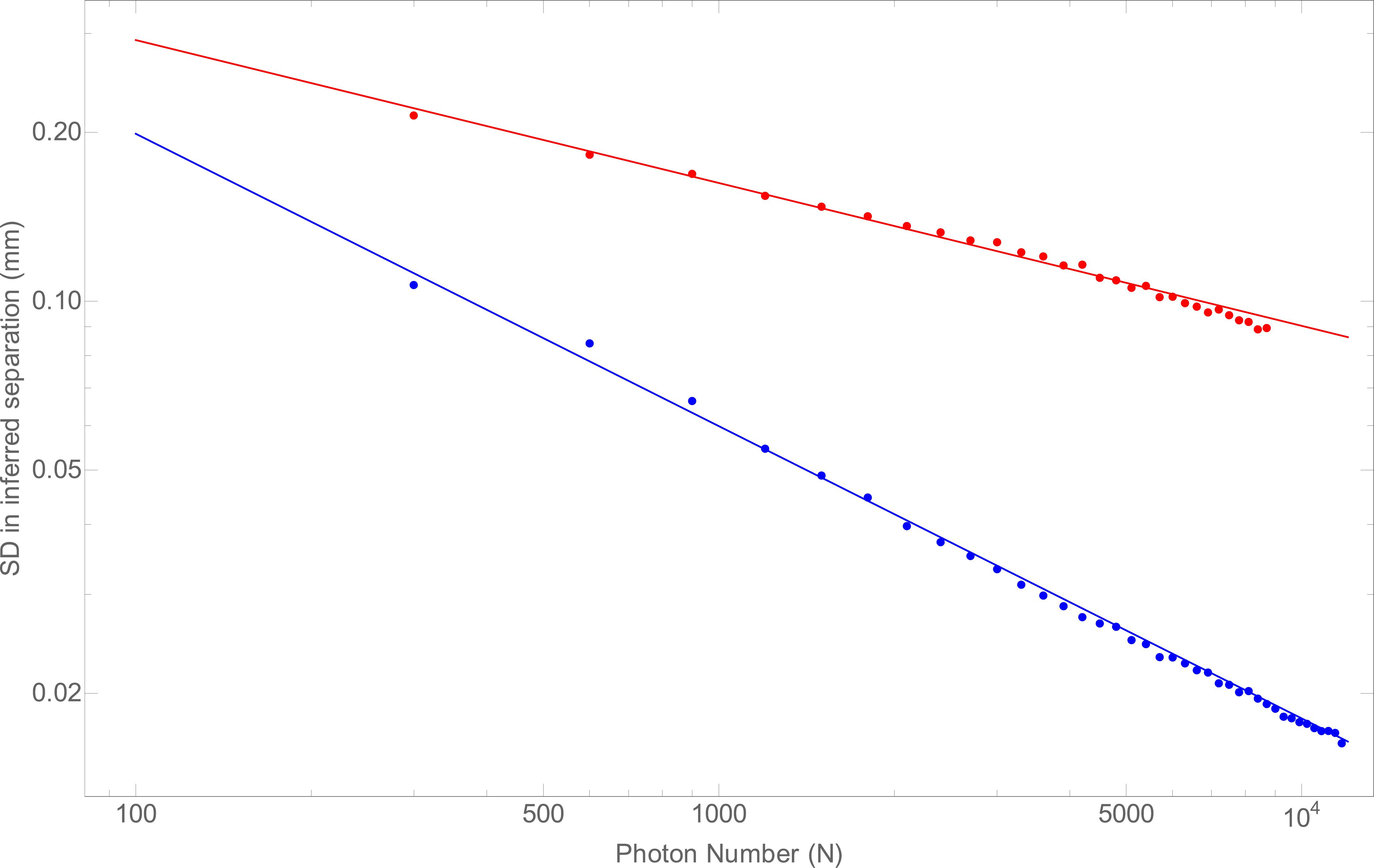}
\par\end{centering}

\caption{\textbf{Monte-Carlo Scaling Analysis. }Dots are the SD in inferred
separation from Monte-Carlo simulations of SPLICE (blue) and IPC (red),
with $\delta=0.21mm$ averaged over 200 repetitions versus total photon
number. Other parameters - i.e. $\sigma$ and $\gamma/N$ (see text)
- were set to experimental values. The overlayed solid lines power-law
fits of the form $\alpha N^{\beta}$ to simulation results. The fit
parameters for IPC are: $\alpha=0.943\pm0.031$, $\beta=-0.254\pm0.004$
and for SPLICE $\alpha=2.188\pm0.090$, $\beta=-0.521\pm0.006$. While
$\delta=0.21mm$ was chosen arbitrarily for this plot, the simulation
was performed for multiple values of $\delta$. We note that with
finite visibility (i.e. $\gamma\protect\neq0$), SPLICE also begins
to scale as $N^{-1/4}$ when both $\delta$ and $N$ are very small.
It is important to note, however, that unlike IPC, this is not a fundamental
scaling but is rather a result of technical limitations (i.e. imperfect
visibility) in our apparatus. A scaling of $N^{-1/2}$ is retained
for all $\delta$ and $N$ when $\gamma=0$.\label{fig:Monte-Carlo-Scaling-Analysis.}}
\end{figure}

\section*{\pagebreak{}}

\section*{Supplementary}

\subsection*{Bias Derivation }

\textbf{\uline{Bias}}

Suppose one was tasked with estimating the value of some parameter,
$x$, by looking at the value of some random variable $y$, which
distribution depends on $x$. If the expectation value of $y$ equals
$x$ for \emph{all }values of the parameter $x$, i.e. $\left\langle y\right\rangle _{x}=x$,
then the random variable $y$ is known as an \emph{unbiased} estimator
of $x$. The distinction between a biased vs an unbiased estimator
is important to consider when one is trying to reason in terms of
the Fisher information and CRLB. As mentioned in the main text, a
more general form of the CRLB is:
\[
\sqrt{\mbox{Var}\left(y\right)}\geq\sqrt{\frac{1}{I_{fish}}\left(1+\frac{\partial\left(\mbox{bias}\right)}{\partial x}\right)}
\]
where bias is $\left\langle y\right\rangle _{x}-x$. Notice that when
the estimator is unbiased, this reduces to equation 4 in the main
text. In the case of IPC, since the Fisher information clearly vanishes
as $\delta\to0$, an infinite variance in the estimator $\delta_{est}$
can be avoided only if\emph{ }the \emph{slope }of $\left\langle y\right\rangle _{x}$
tends to 0 more quickly than $I_{fish}$. It is our goal in this supplementary
section to demonstrate that that is indeed the case.

\textbf{\uline{Spread in estimator of $\delta^{2}$}}

For IPC at small separations each image was fitted to a Taylor expansion
of the detection probability $p_{i}$ (the usual sum of two Gaussians)
to 2nd order:
\[
p_{i}\approx\frac{A}{\sigma\sqrt{2\pi}}\exp\left[-\frac{x_{i}^{2}}{2\sigma^{2}}\right]+\frac{A\delta^{2}}{8\sigma^{5}\sqrt{2\pi}}\exp\left[-\frac{x_{i}^{2}}{2\sigma^{2}}\right]\left(x_{i}^{2}-\sigma^{2}\right)
\]
Subscripts $i$ were added in anticipation of an image consisting
of many pixels at various values of some axis $x$. Performing a linear
regression of a set of photon detection rates $p_{i}$ yields parameters
$A$ and $A\delta^{2}$. Notice that the design matrix, $M$, in this
case contains only $x_{i}$'s and $\sigma$ and so is independent
of photon number $N$. If we now assume that the noise at each pixel
location is mutually independent, then:
\[
\left[\begin{array}{c}
\Delta A\\
\Delta\left(A\delta^{2}\right)
\end{array}\right]=\sqrt{\sum_{j}o_{ij}^{2}\left(\Delta p_{j}\right)^{2}}
\]
where $o_{ij}=\left[\left(M^{T}M\right)^{-1}M^{T}\right]_{ij}$. Supposing
that our \emph{only} source of noise is Poissonian in nature, then
$\Delta p_{j}\sim\sqrt{p_{j}}$ so that $\Delta A$ and $\Delta\left(A\delta^{2}\right)$
both $\sim\sqrt{N}$.

Now elementary error propagation gives:
\[
\Delta\left(\delta^{2}\right)=\sqrt{\left(\frac{\Delta\left(A\delta^{2}\right)}{A}\right)^{2}+\left(\frac{A\delta^{2}}{\Delta A}\right)^{2}}
\]
which in the small $\delta$ limit reduces to
\[
\Delta\left(\delta^{2}\right)\approx\frac{\Delta\left(A\delta^{2}\right)}{A}\sim1/\sqrt{N}
\]
Thus we expect the estimate of $\delta^{2}$ from this method to have
a spread that scales approximately as $1/\sqrt{N}$.

\textbf{\uline{``Clipping'' $\delta^{2}<0$ and bias in $\delta$}}

If there is a sufficiently large number of pixels in our image, the
central limit theorem imposes a Gaussian distribution on $\delta^{2}$,
with width $s=\Delta\left(\delta^{2}\right)=\gamma/\sqrt{N}$ where
$\gamma$ is some constant of proportionality. Although at first glance
negative values of $\delta^{2}$ appear problematic, we can avoid
having to censor parts of our data where this is the case by noting
that they have a natural physical interpretation if we also allow
$\sigma\to-\sigma$ since the quadratic term is paired with an odd
$\sigma^{5}$ term. We can therefore compute the moments of the distribution
relevant to the mean and spread of our estimate of $\delta$:
\begin{eqnarray*}
\left\langle \delta\right\rangle  & = & \frac{1}{s}\sqrt{\frac{2}{\pi}}\int_{-\infty}^{\infty}\sqrt{\left|x\right|}\,\exp\left[-\frac{\left(x-\delta_{actual}^{2}\right)^{2}}{2s^{2}}\right]\,\mbox{d}x\\
\left\langle \delta^{2}\right\rangle  & = & \frac{1}{s}\sqrt{\frac{2}{\pi}}\int_{-\infty}^{\infty}\left|x\right|\,\exp\left[-\frac{\left(x-\delta_{actual}^{2}\right)^{2}}{2s^{2}}\right]\,\mbox{d}x
\end{eqnarray*}
In the limit where $\delta_{actual}\to0$, we find the scalings: $\left\langle \delta\right\rangle \sim\sqrt{s}\sim N^{-1/4}$
and $\left\langle \delta^{2}\right\rangle \sim s\sim N^{-1/2}$, the
latter aggreeing with our Monte Carlo simulations that the standard
deviation or spread in our estimate scales as $N^{-1/4}$ as well.
More crucially, $\left\langle \delta\right\rangle $ can be shown
to approach a constant value sufficiently quickly as $\delta_{actual}$
goes to $0$ for the CRLB to converge to a finite value. $\left\langle \delta\right\rangle $
is plotted in figure 8 of the main text.

Note that the emergence of a bias in our estimate isn't specific to
our treatment of the negative tail of $\delta^{2}$; the same bias
and scalings can be obtained even if we had opted for the lazier approach
of censoring parts of our data that produce negative $\delta^{2}$
values (tantamount to simply ``chopping'' rather than ``folding''
that tail of the distribution). Rather, the bias is more generally
a consequence of performing the regression on $\delta^{2}$ instead
of $\delta$.

The bias vanishes if the two sources have unequal intensities. The
breaking of this symmetry introduces a term in $p_{i}$ that is linear
in $\delta$. If this term is much larger than the quadratic ($\delta^{2}$)
term, we can use $\delta$ as a fit parameter instead, thereby obtaining
an unbiased estimator. We leave the analysis of this asymmetric case
to a possible future work.
\end{document}